# $SU(5)$ grand unified theory with $A_4$ modular symmetry

Francisco J. de Anda,[2,*] Stephen F. King,[1,†] and Elena Perdomo[1,‡]

[1]*School of Physics and Astronomy, University of Southampton, SO17 1BJ Southampton, United Kingdom*
[2]*Tepatitlán's Institute for Theoretical Studies, C.P. 47600, Jalisco, México*

We present the first example of a grand unified theory (GUT) with a modular symmetry interpreted as a family symmetry. The theory is based on supersymmetric $SU(5)$ in 6d, where the two extra dimensions are compactified on a $T_2/\mathbb{Z}_2$ orbifold. We have shown that, if there is a finite modular symmetry, then it can only be $A_4$ with an (infinite) discrete choice of moduli, where we focus on $\tau = \omega = e^{i2\pi/3}$, the unique solution with $|\tau| = 1$. The fields on the branes respect a generalized $CP$ and flavor symmetry $A_4 \ltimes \mathbb{Z}_2$ which is isomorphic to $S_4$ which leads to an effective $\mu - \tau$ reflection symmetry at low energies, implying maximal atmospheric mixing and maximal leptonic $CP$ violation. We construct an explicit model along these lines with two triplet flavons in the bulk, whose vacuum alignments are determined by orbifold boundary conditions, analogous to those used for $SU(5)$ breaking with doublet-triplet splitting. There are two right-handed neutrinos on the branes whose Yukawa couplings are determined by modular weights. The charged lepton and down-type quarks have diagonal and hierarchical Yukawa matrices, with quark mixing due to a hierarchical up-quark Yukawa matrix.

## I. INTRODUCTION

The flavor puzzle, the question of the origin of the three families of quarks and leptons together with their curious pattern of masses and mixings, remains one of the most important unresolved problems of the Standard Model (SM). Following the discovery of neutrino mass and mixing, whose origin is fundamentally unknown, there are now almost 30 undetermined parameters in the SM, far too many for any complete theory. The lepton sector in particular involves large mixing angles that suggest an explanation in terms of discrete non-Abelian family symmetry [1,2]. Furthermore, such discrete non-Abelian family symmetries have been combined with grand unified theories (GUTs) in order to provide a complete description of all quark and lepton (including neutrino) masses and mixings [3].

It is well known that orbifold GUTs in extra dimensions (ED) can provide an elegant explanation of GUT breaking and Higgs doublet-triplet spitting [4]. Similarly, theories involving GUTs and flavor symmetries have been formulated in ED [5–12]. These EDs can help us to understand the origin of the discrete non-Abelian group symmetry such as $A_4$ and $S_4$ which may be identified as a remnant symmetry of the extended Poincaré group after orbifolding.

Some time ago it was suggested that modular symmetry, when interpreted as a family symmetry, might help us to provide a possible explanation for the neutrino mass matrices [13,14]. Recently it has been suggested that neutrino masses might be modular forms [15], with constraints on the Yukawa couplings. This has led to a revival of the idea that modular symmetries are symmetries of the extra-dimensional spacetime with Yukawa couplings determined by their modular weights [16]. However to date, no attempt has been made to combine this idea with orbifold GUTs in order to provide a unified framework for quark and lepton masses and mixings.

In this paper we present the first example in the literature of a GUT with a modular symmetry interpreted as a family symmetry. The theory is based on supersymmetric $SU(5)$ in 6d, where the two extra dimensions are compactified on a $T_2/\mathbb{Z}_2$ orbifold, with a twist angle of $\omega = e^{i2\pi/3}$. Such constructions suggest an underlying modular $A_4$ symmetry with a discrete choice of moduli. This is one of the main differences of the present paper as compared to recent works with modular symmetries which regard the modulus $\tau$ as a free phenomenological parameter [15,16]. We construct a detailed model along these lines where the fields on the branes are assumed to respect a flavor and generalized $CP$ symmetry $A_4 \ltimes \mathbb{Z}_2$ which leads to an effective $\mu - \tau$ reflection symmetry at low energies, implying maximal atmospheric mixing and maximal leptonic $CP$ violation. The model introduces two triplet flavons in the bulk, whose vacuum alignments are determined by orbifold boundary conditions, analogous to those used for $SU(5)$ breaking with doublet-triplet splitting. There are also two right-handed neutrinos on the branes whose Yukawa couplings are determined by modular weights. The charged lepton and down-type quarks have diagonal and hierarchical Yukawa matrices, with quark mixing due to a hierarchical up-quark Yukawa matrix.

[*]fran@tepaits.mx
[†]king@soton.ac.uk
[‡]e.perdomo-mendez@soton.ac.uk

The remainder of the paper is organized as follows. In Sec. II we discuss the orbifold $T^2/\mathbb{Z}_2$ and its symmetries, as follows. In Sec. II A, we give a review of modular transformations while in Sec. II B we describe how the orbifold $T_2/\mathbb{Z}_2$ is only consistent with modular $A_4$ symmetry and a choice of modulus. In Sec. II C, we explicitly show the orbifold $T_2/\mathbb{Z}_2$ with twist angle $\omega = e^{i2\pi/3}$ and modular $A_4$ symmetry. In Sec. II D we study the remnant symmetry after compactification on the $T_2/\mathbb{Z}_2$ orbifold, while Sec. II E connects this remnant symmetry and the modular symmetry. In Sec. II F we discuss the enhanced $A_4 \ltimes \mathbb{Z}_2$ on the branes. In Sec. III, we present the field content of the $SU(5)$ GUT with $A_4$ modular symmetry and a $U(1)$ shaping symmetry, including the Yukawa sector and the specific structure for the effective alignments that the modular symmetry can generate, resulting in the low energy form of the SM fermion mass matrices which we show can lead to a very good fit to the observables. Finally in Sec. IV we present our conclusions. In order to make the paper self-contained, some necessary background information is included in the Appendixes. In Appendix A we show the explicit proof that only the $A_4$ modular symmetry is consistent with the branes, with specific choices of modulus. We supplement the general $A_4$ group theory in Appendix B, the consistency conditions for generalized $CP$ symmetry consistent with $A_4$ in Appendix C and the general theory for modular forms in Appendix D. Finally we show sample fits of the observed data in Appendix E.

## II. ORBIFOLDING AND SYMMETRIES

### A. Review of modular transformations

In this subsection we present the general theory of modular transformations. The structure of the extra-dimensional torus is defined by the structure of the lattice by

$$z = z + \omega_1, \qquad z = z + \omega_2, \tag{1}$$

where $\omega_{1,2}$ are the lattice basis vectors. The variable $z$ refers to the complex coordinate $z = x_5 + ix_6$, where $x_5$ and $x_6$ are the two extra-dimension coordinates. The torus is then characterized by the complex plane $\mathbb{C}$ modulo a two-dimensional lattice $\Lambda_{(\omega_1,\omega_2)}$, where $\Lambda_{(\omega_1,\omega_2)} = \{m\omega_1 + n\omega_2, m, n \in \mathbb{Z}\}$, i.e., $T_2 = \mathbb{C}/\Lambda_{(\omega_1,\omega_2)}$. The lattice is left invariant under a change in lattice basis vectors described by the general transformations

$$\begin{pmatrix} \omega_1 \\ \omega_2 \end{pmatrix} \to \begin{pmatrix} \omega_1' \\ \omega_2' \end{pmatrix} = \begin{pmatrix} a & b \\ c & d \end{pmatrix} \begin{pmatrix} \omega_1 \\ \omega_2 \end{pmatrix},$$

$$\text{where } \begin{pmatrix} a & b \\ c & d \end{pmatrix} \in SL(2, \mathbb{Z}) \tag{2}$$

or equivalently if $a, b, c, d \in \mathbb{Z}$ and $ad - bc = 1$. These are called modular transformations and form the modular group $\Gamma$ [15]. Without loss of generality, the lattice vectors may be rescaled as

$$\begin{pmatrix} \omega_1 \\ \omega_2 \end{pmatrix} \to \begin{pmatrix} 1 \\ \tau \end{pmatrix} = \begin{pmatrix} 1 \\ \omega_2/\omega_1 \end{pmatrix}, \tag{3}$$

such that the torus is equivalent to one whose periods are 1 and $\tau = \omega_2/\omega_1$ and we can restrict $\tau$ to the upper half-plane $\mathcal{H} = \text{Im}\tau > 0$. The modular transformations on the rescaled basis vectors which leave the lattice invariant are given by[1]

$$\begin{pmatrix} 1 \\ \tau \end{pmatrix} \to \begin{pmatrix} 1 \\ \tau' \end{pmatrix}, \quad \text{where } \tau' = \frac{\omega_2'}{\omega_1'} = \frac{a\tau + b}{c\tau + d}. \tag{4}$$

A $SL(2, \mathbb{Z})$ transformation on the modulus parameter $\tau$ and its negative are equivalent, as can be seen from Eqs. (2) and (4). Therefore, we can use the infinite discrete group $PSL(2, \mathbb{Z}) = SL(2, \mathbb{Z})/\mathbb{Z}_2$, generated by

$$S\colon \tau \to -1/\tau \quad \text{and} \quad T\colon \tau \to \tau + 1, \tag{5}$$

to describe the transformations that relate equivalent tori. This is also called the modular group $\bar{\Gamma}$ satisfying $\bar{\Gamma} = \Gamma/\{\pm 1\}$.[2] The generators of the infinite-dimensional modular group can also be written as

$$S = \begin{pmatrix} 0 & 1 \\ -1 & 0 \end{pmatrix}, \quad T = \begin{pmatrix} 1 & 0 \\ 1 & 1 \end{pmatrix}. \tag{6}$$

They satisfy the presentation

$$\bar{\Gamma} \simeq \{S, T | S^2 = (ST)^3 = \mathbb{I}\}/\{\pm 1\}, \tag{7}$$

where $S, T \in SL(2, \mathbb{Z})$.

We will be considering the finite-dimensional discrete subgroups by imposing an additional constraint on $T^M$, where $M$ is a positive integer,

$$\bar{\Gamma}_M \simeq \{S, T | S^2 = (ST)^3 = T^M = \mathbb{I}\}/\{\pm 1\}, \tag{8}$$

---

[1]Where we have relabeled the integers $a \to d$, $b \to c$, $c \to b$, $d \to a$ to be consistent with conventional notation.
[2]The modular group $\Gamma$ refers to $SL(2, \mathbb{Z})$, while $\bar{\Gamma}$ is used for $PSL(2, \mathbb{Z})$ and takes into account the equivalence of an $SL(2, \mathbb{Z})$ matrix and its negative.

where $S, T \in SL(2, \mathbb{Z}_M)$. These groups, with $M \leq 5$, are isomorphic to the known discrete groups as $\bar{\Gamma}_2 \simeq S_3$, $\bar{\Gamma}_3 \simeq A_4$, $\bar{\Gamma}_4 \simeq S_4$, $\bar{\Gamma}_5 \simeq A_5$.

We now introduce a convenient (if nonunique) representation for the modular transformations consistent with the presentation in Eq. (8),

$$S = \begin{pmatrix} 0 & 1 \\ -1 & 0 \end{pmatrix}, \quad T_{(M)} = \begin{pmatrix} e^{-2i\pi/M} & 0 \\ 1 & e^{2i\pi/M} \end{pmatrix}, \quad (9)$$

which satisfies the presentation of the $\bar{\Gamma}_M$ group, for any integer $M > 2$. This representation will be useful in the following discussion.

### B. Why the orbifold $T^2/\mathbb{Z}_2$ suggests modular $A_4$ symmetry with modulus $\tau = \omega$

In this subsection we present an argument which shows that a particular $T^2/\mathbb{Z}_2$ orbifold (as assumed in this paper) suggests an underlying modular $A_4$ symmetry with specific modulus parameters.

We begin by defining the orbifold $T^2/\mathbb{Z}_2$ in terms of two arbitrary lattice vectors $\omega_1$ and $\omega_2$,

$$z = z + \omega_1, \quad z = z + \omega_2, \quad z = -z. \quad (10)$$

The action of the orbifold in Eq. (10) leaves four invariant 4d branes given by[3]

$$\bar{z} = \left\{ 0, \frac{\omega_1}{2}, \frac{\omega_2}{2}, \frac{\omega_1 + \omega_2}{2} \right\}. \quad (11)$$

After compactification, the symmetries of the branes remain unbroken; therefore it is relevant to study any possible symmetry among the branes which will affect the fields localized on them. Therefore, we want to check if the modular transformations in Eq. (9) leave an invariant set of branes for some value of $M$. At this stage the modulus $\tau = \omega_2/\omega_1$ can apparently take any value. However we present a proof in Appendix A that only the $A_4$ symmetry is consistent, meaning that $M = 3$, when the basis vectors are related by

$$\omega_2 = -\frac{e^{i2\pi/3} + 2p + 2}{2q + 1} \omega_1. \quad (12)$$

The $p$ and $q$ are integers satisfying that

$$r = \frac{(2p+1)(p+1) + q + 1}{2q+1} \quad (13)$$

is an integer, which has infinitely many discrete solutions. Furthermore, since the modular forms restrict $\tau$ to be in the upper complex plane, then $q < 0$.

---

[3]The notation for the lattice vectors $\omega_{1,2}$ should not be confused with the twist angle $\omega = e^{i2\pi/3}$.

This paper's approach is to focus on the orbifold first and then derive the modular symmetries, instead of going directly into the modular symmetries. We will restrict ourselves to the case where $|\omega_1| = |\omega_2|$, which happens when $p = -1$, $q = -1$. This way we focus on studying the effects only of the angle between both vectors. We can, without loss of generality, choose $\omega_1 = 1$. Furthermore, the modular symmetries require $\tau$ to lie in the upper complex plane; in this case the only solutions to Eqs. (12) and (13) are $\omega_2 = \omega = e^{i2\pi/3}$. This uniquely fixes the modulus coming from the orbifold $T^2/\mathbb{Z}_2$. We emphasize that this is one of the main differences of the present paper as compared to recent works with modular symmetries which regard the modulus $\tau$ as a free phenomenological parameter [15,16]. In our work, we assume a specific orbifold $T^2/\mathbb{Z}_2$, for which we have shown that one consistent choice for a surviving modular symmetry is $A_4$ with fixed modulus $\tau$, although we shall not address the problem of moduli stabilization [17].

### C. The orbifold $T^2/\mathbb{Z}_2$ with $\omega = e^{i2\pi/3}$ and modular $A_4$ symmetry

Following the argument of the previous subsection, we henceforth focus on the orbifold $T^2/\mathbb{Z}_2$ with particular twist angle denoted as $\omega = e^{i2\pi/3}$, identified as the modulus $\tau$ associated with a particular finite modular symmetry $A_4$, where $A_4$ is the only choice consistent with this orbifold.

This orbifold then corresponds to the identification

$$z = z + 1, \quad z = z + \omega, \quad z = -z, \quad (14)$$

where the first two equations are the periodic conditions from the torus $T^2$ and the third one is the action generated by the orbifolding symmetry $\mathbb{Z}_2$. The twist corresponds to $\omega = e^{i2\pi/3}$. The orbifold symmetry transformations leave four invariant 4d branes as shown in Fig. 1:

$$\bar{z} = \left\{ 0, \frac{1}{2}, \frac{\omega}{2}, \frac{1+\omega}{2} \right\}. \quad (15)$$

The transformations

$$S: z \to z + 1/2 \quad \text{or} \quad z \to z + \omega/2,$$
$$T: z \to \omega^2 z,$$
$$U: z \to z^* \quad \text{or} \quad z \to -z^*, \quad (16)$$

permute the branes and leave invariant the set of four branes in Eq. (15). These transformations satisfy

$$S^2 = T^3 = (ST)^3 = 1,$$
$$U^2 = (SU)^2 = (TU)^2 = (STU)^4 = 1, \quad (17)$$

where the first line is the presentation of the group $A_4$ and both lines complete the presentation of $S_4$ [1]. In Fig. 1 we show how these transformations act on the extra-

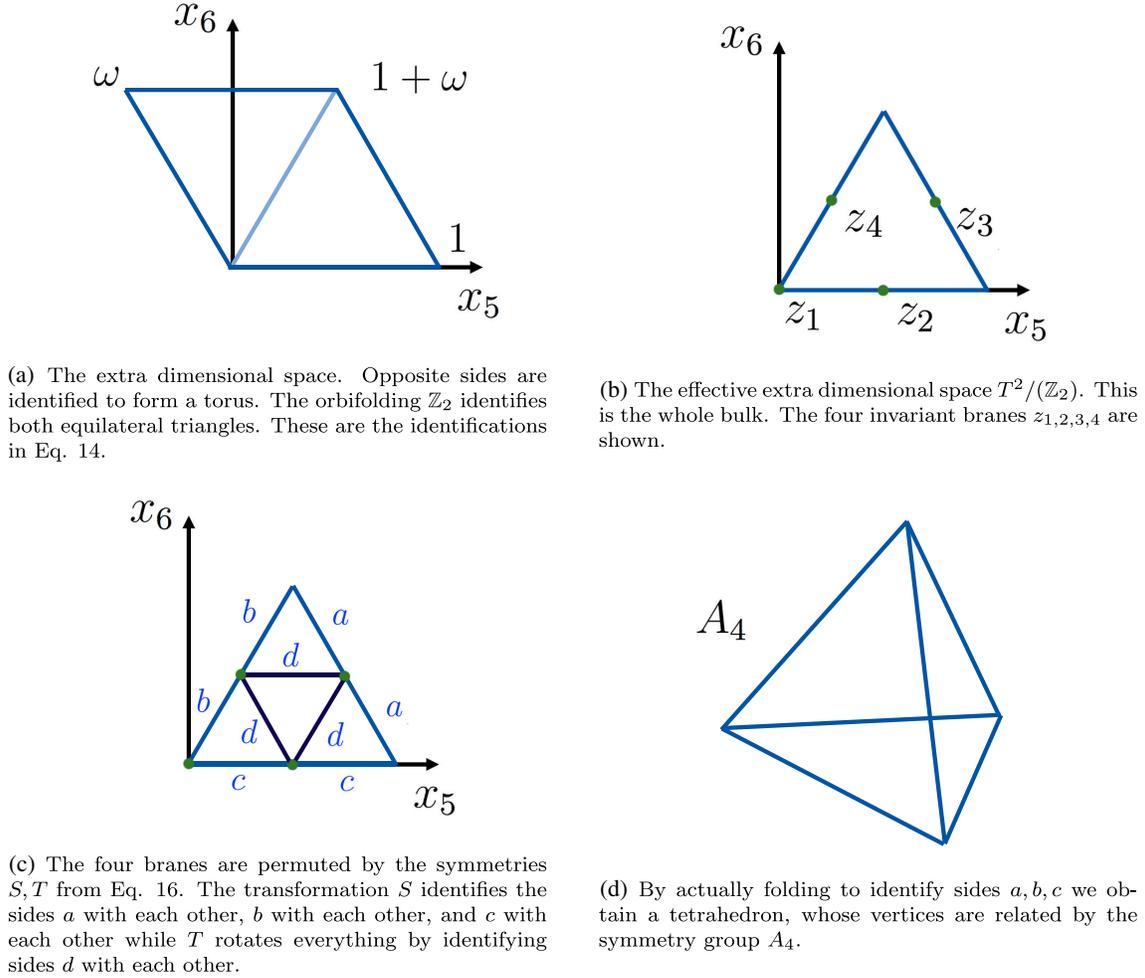

(a) The extra dimensional space. Opposite sides are identified to form a torus. The orbifolding $\mathbb{Z}_2$ identifies both equilateral triangles. These are the identifications in Eq. 14.

(b) The effective extra dimensional space $T^2/(\mathbb{Z}_2)$. This is the whole bulk. The four invariant branes $z_{1,2,3,4}$ are shown.

(c) The four branes are permuted by the symmetries $S, T$ from Eq. 16. The transformation $S$ identifies the sides $a$ with each other, $b$ with each other, and $c$ with each other while $T$ rotates everything by identifying sides $d$ with each other.

(d) By actually folding to identify sides $a, b, c$ we obtain a tetrahedron, whose vertices are related by the symmetry group $A_4$.

FIG. 1. Visualization on the remnant $A_4$ symmetry after orbifolding.

dimensional space and how the "remnant $A_4$ symmetry" is realized.

Fixing $M = 3$, the set of branes is invariant under the modular transformations

$$S = \begin{pmatrix} 0 & 1 \\ -1 & 0 \end{pmatrix}, \qquad T_{(3)} = \begin{pmatrix} \omega^2 & 0 \\ 1 & \omega \end{pmatrix}, \qquad (18)$$

on the lattice vectors $(1, \omega)^T$. These transform the basis vectors as

$$S\begin{pmatrix} 1 \\ \omega \end{pmatrix} = \begin{pmatrix} \omega \\ -1 \end{pmatrix}, \quad T_{(3)}\begin{pmatrix} 1 \\ \omega \end{pmatrix} = \begin{pmatrix} \omega^2 \\ 1+\omega^2 \end{pmatrix} = \begin{pmatrix} -1-\omega \\ -\omega \end{pmatrix}, \qquad (19)$$

(noting that $1 + \omega + \omega^2 = 0$) leaving the lattice invariant as can be seen from Fig. 2.

The matrices $S, T_{(3)}$ fulfill the presentation of the group they generate to be

$$\{S, T_{(3)} | S^2 = T_{(3)}^3 = (ST_{(3)})^3 = \mathbb{I}\}/\{\pm 1\} = \bar{\Gamma}_3 \simeq A_4, \quad (20)$$

where $S, T_{(3)} \in SL(2, \mathbb{Z}_3)$, so that the branes are indeed invariant under the discrete modular group $\bar{\Gamma}_3 \simeq A_4$. As we will see in Sec. II F, this symmetry will be enlarged.

### D. Remnant brane symmetry for $T^2/\mathbb{Z}_2$ with $\omega = e^{i2\pi/3}$

So far we have shown that the choice of orbifold $T^2/\mathbb{Z}_2$ is consistent with the finite modular symmetry $A_4$ with a discrete choice of moduli, where we focus on $\tau = \omega = e^{i2\pi/3}$. Now we will take a step back, forget about modular symmetries for a while, and just consider the symmetries of the branes with a twist angle $\omega = e^{i2\pi/3}$. We will discover an $S_4$ symmetry that has apparently nothing to do with modular symmetry, which we refer to as "remnant $S_4$ symmetry." In the next subsection we shall show how the subgroup remnant $A_4$ symmetry is related to the previous $A_4$ finite modular symmetry.

In this section, we will find that the branes are invariant under an $S_4$ and its subgroup $A_4$ symmetry which can be identified as a remnant symmetry of the spacetime symmetry after it is broken down to the 4d Poincaré symmetry

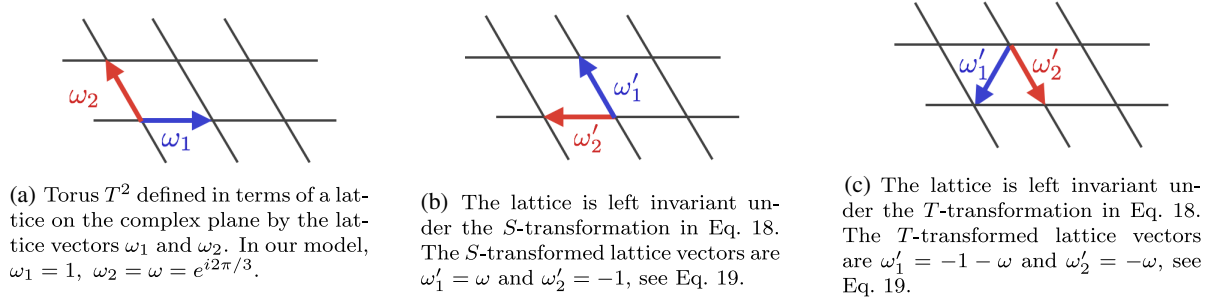

(a) Torus $T^2$ defined in terms of a lattice on the complex plane by the lattice vectors $\omega_1$ and $\omega_2$. In our model, $\omega_1 = 1$, $\omega_2 = \omega = e^{i2\pi/3}$.

(b) The lattice is left invariant under the $S$-transformation in Eq. 18. The $S$-transformed lattice vectors are $\omega_1' = \omega$ and $\omega_2' = -1$, see Eq. 19.

(c) The lattice is left invariant under the $T$-transformation in Eq. 18. The $T$-transformed lattice vectors are $\omega_1' = -1 - \omega$ and $\omega_2' = -\omega$, see Eq. 19.

FIG. 2. Visualization of the lattice invariant transformations, $S$ and $T$.

through orbifold compactification. Here, we assume that the spacetime symmetry before compactification is a 6d Poincaré symmetry. The compactification breaks part of this symmetry. However, due to the geometry of our orbifold with twist angle $\omega = e^{i2\pi/3}$, a discrete subgroup is left unbroken. This group may be generated by the spacetime transformations (which belong to the extra-dimensional part of the 6d Poincaré).

The orbifolding leaves four invariant branes, and this specific orbifold structure leaves them related by the group $S_4$. This symmetry, together with 4d Poincaré transformations, is a subgroup of the extra-dimensional Poincaré symmetry that survives compactification. This is the standard "remnant symmetry" [6,18].

Any field located in the branes will transform under the 4d Poincaré group as usual. Since the branes transform into each other by the remnant symmetries, the fields on the brane should also transform under them. The four branes transform under the remnant $S_4$ symmetry and we choose the embedding of the representation $4 \to 3 + 1$ so that the fields in the branes can only transform under those irreducible representations [7,11]. The fields in the branes do not depend on $z$ but are permuted into each other by the $S$, $T$, $U$ transformations.

### E. The connection between remnant $A_4$ symmetry and finite modular $A_4$ symmetry

We have shown that the set of branes is invariant under a remnant $S_4$ (and its subgroup $A_4$) subgroup of the extra-dimensional Poincaré symmetry. We shall return to the $S_4$ symmetry in the next subsection and we now show that the remnant $A_4$ symmetry can be identified with the finite modular $A_4$ symmetry discussed earlier. Essentially, if we impose a modular symmetry $A_4$ on the whole space, its action on the branes is the same action as the remnant spacetime symmetry; i.e., it permutes the branes but leaves invariant the whole set. The modular symmetry acts on the basis vectors of the torus while the remnant symmetry is a spacetime symmetry and acts on the fixed points; therefore from the point of view of the branes, the "remnant $A_4$ symmetry" is an active transformation while the finite modular $A_4$ symmetry is an equivalent passive transformation. This way we may identify the remnant $A_4$ symmetry of the branes as a modular symmetry since the effect on the branes of each type of transformation is identical; it is just a choice of "picture" (active or passive) which we choose.

The action of the modular symmetry on the branes behaves as a "normal" symmetry (i.e., modular forms are not relevant) since fields located on the brane do not depend on the extra-dimensional coordinate. The modular symmetry can therefore be imposed as any usual symmetry. In the orbifold $T_2/\mathbb{Z}_2$, the branes are only consistent with the modular group $\bar{\Gamma}_3$, as shown in Appendix A. Any theory with this orbifold and fields allocated on the branes will only be consistent with the modular symmetry $\bar{\Gamma}_3$. In such a setup, the branes see the finite modular symmetry as simply equivalent to a remnant symmetry, a subgroup of the extra-dimensional Poincaré group.

We can see in Eq. (18) that the $S$, $T$ transformations (and therefore the $\bar{\Gamma}_3$ modular transformations) correspond to specific passive reflections, rotations and translations. In this way the $\bar{\Gamma}_3$ must be a subgroup of the 6d Poincaré group. All modular groups are. However not all modular groups are consistent with the invariant branes, as we have shown.

On the other hand, fields in the bulk, which feel the extra dimensions, will also transform under some representation of the 6d Poincaré; however in this case they will transform under a nonlinear realization of this $\bar{\Gamma}_3$ symmetry, and this is precisely what are referred to as the modular forms [15].

We conclude that the modular symmetry $\bar{\Gamma}_3$ acts either as a linear or nonlinear realization of the remnant symmetry $A_4$, depending on whether we are concerned with brane fields or bulk fields.

### F. Enhanced $A_4 \ltimes \mathbb{Z}_2$ symmetry of the branes

We now recall that, in our setup, the brane fields enjoy a larger $S_4$ symmetry than the remnant $A_4$ symmetry, as shown in Sec. II D. However this larger $S_4$ symmetry is not related to a finite modular symmetry, since the branes can only be invariant under the modular transformations corresponding to $\bar{\Gamma}_3 \simeq A_4$, and is not enjoyed by the fields in the bulk.

We note here that $S_4 \simeq A_4 \ltimes \mathbb{Z}_2$. The symmetry generated by $U$ from Eq. (16) is a remnant symmetry of the orbifolding process, but it cannot be interpreted as a modular transformation. We conclude that the remnant symmetry of the branes is $\bar{\Gamma}_3 \ltimes \mathbb{Z}_2 \simeq A_4 \ltimes \mathbb{Z}_2$. The $\mathbb{Z}_2$ symmetry is generated by $C \cdot U$ where $U$ is the usual matrix representation of the generator from $S_4$ and $C$ stands for complex conjugation of the complex coordinate, which is equivalent to a change of sign in $x_6$, i.e., the parity transformation of the sixth dimension $P_6$. The $\mathbb{Z}_2$ is not a modular symmetry while the $A_4$ is. The product of both symmetries is not direct since the generator $U$ does not commute with all $A_4$ generators and is the corresponding $S_4$ generator.

After compactification, the remnant $\mathbb{Z}_2$ (which is not a subset of $A_4$) acts on the brane fields generalized $CP$ transformation where the transformations $P_1, ..., P_5$ are trivial while $\tilde{P}_6 = P_6 U$, where $P_6$ is the trivial parity transformation, while the $U$ is a family transformation [19]; thus a field transforms as

$$P_i \phi(x_1, ..., x_i, ..., x_6) = \phi(x_1, ..., -x_i, ..., x_6),$$
$$\text{for } i = 1, ...5,$$
$$P_6 \phi(x_1, ..., x_i, ..., x_6) = U \phi(x_1, ..., x_i, ..., -x_6). \quad (21)$$

Under $CP$ the fields transform as shown in Appendix C. As stated before, this effective symmetry transformation only affects nontrivially the brane fields and the fields in the bulk are unaffected, transforming under canonical $CP$ and not forced to preserve it. Thus in our approach the generalized $CP$ is a remnant symmetry in a particular sector of the theory, corresponding to the fields on the branes.

We have shown that the remnant $\mathbb{Z}_2$ symmetry on the branes behaves as an effective generalized $CP$ transformation. In Appendix C we check its compatibility with the $A_4$ flavor symmetry and find that it is consistent, as indeed it must be.

### III. $SU(5)$ GUT WITH $A_4$ MODULAR SYMMETRY

#### A. The model

In this section we construct a supersymmetric $SU(5)$ GUT model on a 6d orbifold $T^2/\mathbb{Z}_2$ with twist $\omega = e^{i2\pi/3}$, with an $A_4$ modular symmetry as a flavor symmetry, extended by the $\mathbb{Z}_2$ symmetry on the branes. Furthermore we impose a global $U(1)$ as a shaping symmetry. We impose different boundary conditions at each invariant brane, which break the original symmetry into the minimal supersymmetric standard model (MSSM). The $U(1)$ as a shaping symmetry forbids any higher order terms, while a discrete $\mathbb{Z}_N$ would allow them, the smaller the $N$, the corrections would appear at lower order.

The $A_4$ modular symmetry will require the Yukawa couplings to be specific modular forms, while the $\mathbb{Z}_2$ symmetry will further restrict the possible mass matrix structure so that the theory has strong predictions for leptons [20]. As we shall see later, the up quarks will lie in different $A_4$ singlets with modular weight zero, so that only the subgroup $\mathbb{Z}_3$ is a remnant while the $\mathbb{Z}_2$ behaves trivially. This forces stringent relations for the lepton mass matrices but not for the quarks.

All the fields in the bulk $\psi$ will transform under the modular transformations

$$\tau \to \frac{a\tau + b}{c\tau + d}, \qquad \psi \to (c\tau + d)^{-k} \rho \psi, \quad (22)$$

where $\rho$ is the usual matrix representation of the corresponding $A_4$ transformation. Each field has a weight $-k$, with no constraint in $k$ since the fields are not modular forms. The superfields that are located on the brane do not depend on the extra dimensions and therefore they must have weight zero [15]. We arbitrarily choose a weight for each of the bulk fields.

The whole field content is listed in Tables I and II. The fields that do not have weight or parity under the boundary conditions are located on the branes and feel the symmetry $A_4 \ltimes \mathbb{Z}_2$; see Table I. The transformations of the fields under this symmetry are discussed in Appendix B. The **3** representations on the brane transform under $A_4 \ltimes \mathbb{Z}_2$ are as shown in Table VI and Eq. (C4).

TABLE I. Fields on the branes, including matter and right-handed neutrino superfields. A working set of charges is $\{q_1, q_2, q_3\} = \{2, 0, 1\}$. Note that the **3** representations on the brane transform under $A_4 \ltimes \mathbb{Z}_2$ as shown in Table VI and Eq. (C4).

|  | Representation | | |
|---|---|---|---|
| Field | $A_4 \ltimes \mathbb{Z}_2$ | $SU(5)$ | $U(1)$ |
| $F$ | **3** | $\bar{\mathbf{5}}$ | $q_1 + 2q_3$ |
| $N^c_s$ | **1** | **1** | $q_1$ |
| $N^c_a$ | **1** | **1** | $4q_1$ |
| $\xi$ | **1** | **1** | $-2q_1$ |

TABLE II. Fields on the bulk used in constructing the model, including matter, Higgs and flavon superfields. A working set of charges is $\{q_1, q_2, q_3\} = \{2, 0, 1\}$.

|  | Representation | | | | Localization | | |
|---|---|---|---|---|---|---|---|
| Field | $A_4$ | $SU(5)$ | $U(1)$ | Weight | $P_0$ | $P_{1/2}$ | $P_{\omega/2}$ |
| $T_1^\pm$ | $\mathbf{1}''$ | **10** | $q_3 + 4q_1$ | $-\gamma$ | $+1$ | $\pm 1$ | $\pm 1$ |
| $T_2^\pm$ | $\mathbf{1}'$ | **10** | $q_3 + 2q_1$ | $-\gamma$ | $+1$ | $\pm 1$ | $\pm 1$ |
| $T_3^\pm$ | **1** | **10** | $q_3$ | $-\gamma$ | $+1$ | $\pm 1$ | $\pm 1$ |
| $H_5$ | **1** | **5** | $-2q_3$ | $-\alpha$ | $+1$ | $+1$ | $+1$ |
| $H_{\bar{5}}$ | $\mathbf{1}'$ | $\bar{\mathbf{5}}$ | $q_2$ | $\alpha + \gamma$ | $+1$ | $+1$ | $+1$ |
| $\phi_1$ | **3** | **1** | $-q_2 - q_1 - 3q_3$ | $-\alpha$ | $+1$ | $+1$ | $+1$ |
| $\phi_2$ | **3** | **1** | $-3q_1$ | $\alpha - \beta$ | $+1$ | $-1$ | $+1$ |

The field $F$ contains the MSSM fields $L$ and $d_R$ and is a flavor triplet. It is located on the brane. The fields $T_i^{\pm}$ contain the MSSM $u_R, e_R, Q$; they are three flavor singlets. There are two copies of each $T$ with different parities under the boundary conditions; as we shall see in the next section, this allows different masses for down quarks and charged leptons. There are only two right-handed neutrinos $N_{a,s}^c$. The MSSM Higgs fields $h_{u,d}$ are inside the $H_{5,\bar{5}}$ respectively. We have two flavons $\phi_{1,2}$ that help to give structure to the fermion masses. Finally, the field $\xi$ generates the hierarchy between the masses à la Froggat-Nielsen [21].

### B. GUT and flavor breaking by orbifolding

Since the orbifold has the symmetry transformations of Eq. (14), the fields must also comply with them. However since we are in a gauge theory, the equations need not be fulfilled exactly but only up to a gauge transformation, so any field complies with

$$\phi(x, z) = G \, \phi(x, -z),$$
$$\phi(x, z) = G_5 \, \phi(x, z + 1),$$
$$\phi(x, z) = G_6 \, \phi(x, z + \omega), \quad (23)$$

where the $G$'s are gauge transformations that must fulfill

$$G^2 = 1, \qquad G_5 G_6 = G_6 G_5, \qquad G G_{5,6} G = G_{5,6}^{-1}, \quad (24)$$

where the first equation comes from the fact that it belongs to the parity operator, the second is due to the fact of the commutativity of the translations and the third one denotes the relation between parity and translations.

Since the branes $\bar{z}_i$ are invariant under the orbifold symmetry transformations of Eq. (14), they act as boundaries which, due to the $G$'s gauge transformations, impose the boundary conditions

$$\phi(x, z + \bar{z}_i) = P_{\bar{z}_i} \phi(x, -z + \bar{z}_i), \quad (25)$$

which correspond to a reflection at each of the branes. These boundary conditions are related to the gauge transformations as

$$P_0 = G, \qquad P_{1/2} = G_5 G,$$
$$P_{\omega/2} = G_6 G, \qquad P_{(1+\omega)/2} = G_5 G_6 G. \quad (26)$$

For simplicity, we choose all $G$'s to commute, meaning that $G_{5,6} = G_{5,6}^{-1}$, and therefore all boundary conditions become matrices of order 2. The boundary conditions imply an invariance at each brane under some $A_4 \times SU(5)$ transformation, and they are chosen to break the symmetry in a particular way as follows:

$$P_0 = \mathbb{I}_3 \times \mathbb{I}_5,$$
$$P_{1/2} = T_1 \times \text{diag}(-1, -1, -1, 1, 1),$$
$$P_{\omega/2} = T_2 \times \text{diag}(-1, -1, -1, 1, 1), \quad (27)$$

where $\mathbb{I}_3, T_{1,2} \in SU(3)$, while $\mathbb{I}_5, \text{diag}(-1, -1, -1, 1, 1) \in SU(5)$, and explicitly

$$T_1 = \begin{pmatrix} 1 & 0 & 0 \\ 0 & -1 & 0 \\ 0 & 0 & -1 \end{pmatrix}, \quad T_2 = \begin{pmatrix} 1 & 0 & 0 \\ 0 & 0 & 1 \\ 0 & 1 & 0 \end{pmatrix} = U, \quad (28)$$

and the last boundary condition is fixed and defined by the other boundary conditions as $P_{(1+\omega)/2} = P_0 P_{1/2} P_0 P_{\omega/2} P_0 = T_1 T_2 \times \mathbb{I}_5$.

The boundary condition $P_0$ breaks the effective extended $\mathcal{N} = 2 \to \mathcal{N} = 1$ SUSY. The boundary conditions $P_{1/2, \omega/2}$ leave their corresponding $\mathbb{Z}_2$ symmetry invariant and together break $A_4$ completely and $SU(5) \to SU(3) \times SU(2) \times U(1)$.

The fields $F, N_{a,s}^c, \xi$ lie on the brane and are unaffected by the boundary conditions. The fields $T^{\pm}$ are $A_4$ singlets and do not feel the $A_4$ breaking conditions. They have different parities and feel the $SU(5)$ breaking condition. The fields $T^+$ contain the light MSSM $u_R, e_R$ fields, while $T^-$ contains the light field $Q$. This allows for independent masses for charged leptons and down quarks since they come from different fields. The Higgs fields feel the $SU(5)$ breaking condition leaving only the light doublets, solving the doublet triplet splitting problem [7] (for a recent discussion see e.g., [11]).

The flavons $\phi_{1,2}$ feel the $A_4$ breaking conditions. They have different parities under the conditions and this fixes their alignments to be

$$\langle \phi_1 \rangle = v_1 \begin{pmatrix} 1 \\ 0 \\ 0 \end{pmatrix}, \qquad \langle \phi_2 \rangle = v_2 \begin{pmatrix} 0 \\ 1 \\ 1 \end{pmatrix}. \quad (29)$$

We may remark that these flavon vacuum expectation value (VEV) alignments do not break the $\mathbb{Z}_2$ symmetry generated by $U$, even though they are in the bulk.

We see that the orbifolding breaks the symmetry $SU(5) \times A_4 \ltimes \mathbb{Z}_2 \to SU(3) \times SU(2) \times U(1) \times \mathbb{Z}_2$ while solving the doublet triplet splitting, separating charged lepton and down-quark masses and completely aligning flavon VEVs.

We do not show an explicit driving mechanism for the VEVs $v_{1,2,\xi}$. We assume that they are driven radiatively [22].

### C. Yukawa structure

In 6d, the superpotential has dimension 5 while each superfield has dimension 2. A 6d interacting superpotential is inherently nonrenormalizable. We work with the effective 4d superpotential, which happens after compactification. We assume the compactification scale is close to the original cutoff scale. We use $\Lambda$ to denote both the compactification scale and the GUT scale, which is taken to be the cutoff of the effective theory. Assuming this makes the Kaluza-Klein modes to be at the GUT scale so that they do not spoil standard gauge coupling unification or any of the current precision tests.

With the fields in Tables I and II, we can write the effective 4d Yukawa terms

$$\begin{aligned} \mathcal{W}_Y = & y_s^N \xi N_s^c N_s^c + y_a^N \xi \frac{\xi^3}{\Lambda^3} N_a^c N_a^c \\ & + y_s^\nu \frac{\xi}{\Lambda} F H_5 N_s^c + y_a^\nu \frac{\phi_2 \xi}{\Lambda^2} F H_5 N_a^c \\ & + y_3^e \frac{\phi_1}{\Lambda} F H_{\bar{5}} T_3^+ + y_2^e \frac{\phi_1 \xi}{\Lambda^2} F H_{\bar{5}} T_2^+ + y_1^e \frac{\phi_1 \xi^2}{\Lambda^3} F H_{\bar{5}} T_1^+ \\ & + y_3^d \frac{\phi_1}{\Lambda} F H_{\bar{5}} T_3^- + y_2^d \frac{\phi_1 \xi}{\Lambda^2} F H_{\bar{5}} T_2^- + y_1^d \frac{\phi_1 \xi^2}{\Lambda^3} F H_{\bar{5}} T_1^- \\ & + y_{ij}^u H_5 T_i^+ T_j^- \frac{\xi^{6-i-j}}{\Lambda^{6-i-j}}, \end{aligned} \quad (30)$$

where $i, j = 1, 2, 3$. Due to the stringent $U(1)$ shaping symmetry, there are no higher order terms. The field $\xi$ has a VEV and generates hierarchies between families à la Froggatt-Nielsen [21].

The first line in Eq. (30) gives the two right-handed (RH) neutrino Majorana masses without any mixing. The fields in both terms have zero weight so the modular symmetry does not add anything new. The second line generates Dirac neutrino masses. They have nontrivial weights and their structure will be discussed in Sec. III D. The third line gives masses to charged leptons. They are all weight zero automatically and the mass matrix is diagonal. The fourth line generates a diagonal down-quark mass matrix. Since it involves a different field ($T^-$ instead of $T^+$) the coupling constants are independent. Finally the fifth line gives masses to the up quarks. It is a general nonsymmetric mass matrix with complex entries. Since the fields in these terms have a nontrivial weight but the $T^\pm$ are singlets, the modular symmetry does not change the matrix structure. We remark that the top-quark mass term is renormalizable.

At the GUT level, the $\mu$ term is forbidden, so it should be generated by another mechanism at a much smaller scale [23].

### D. Effective alignments from modular forms

In Eq. (30) we have a few terms involving nontrivial weights under the modular symmetry. This implies that the couplings

$$y_s^\nu, \quad y_a^\nu, \quad y_{ij}^u \quad (31)$$

are modular forms with a positive even weight [24]. They involve the Dedekind $\eta$ function and its exact form can be found in Appendix D.

The modular forms are functions of the lattice basis vector parameter $\tau$ from Eq. (3). Usually this parameter is chosen to give a good fit to the flavor parameters. In our case, the specific orbifold of our model is set to fix

$$\tau = \omega = e^{2i\pi/3}, \quad (32)$$

and the modular form structure is fixed up to a real constant due to the extra condition coming from the generalized $CP$ symmetry.

The modular form $y_s^\nu$ must be a triplet under $A_4$ to construct an invariant singlet with the triplet field $F$. Furthermore, it has weight $\alpha$ to compensate the overall weight of the corresponding term. We show the effective triplet alignments it can have in Table III for different weights $\alpha$. The possibilities are very limited since many modular forms vanish when $\tau = \omega$, as shown in Appendix D. Larger weight modular forms repeat the same structure so that this table is exhaustive, as discussed in Appendix D.

The modular form $y_a^\nu$ must have weight $\beta$. It multiplies the flavon $\phi_2$, so that they must be contracted into a triplet $(y_a^\nu \langle \phi_2 \rangle)_3$ which will generate the effective alignment. In the case of $y_a^\nu$ being a singlet under $A_4$, the effective alignment is simply given the flavon VEV $\langle \phi_2 \rangle$ in Eq. (29), which was fixed by the orbifold boundary conditions. When $y_a^\nu$ is a triplet under $A_4$, it must be contracted with $\phi_2$ as shown in Appendix B, $3 \times 3 \to 1 + 1' + 1'' + 3_a + 3_s$. This gives different possible products for the effective triplet. The actual effective alignment is an arbitrary linear combination of all possibilities and can be found in Table IV. For $\beta = 0$ the only modular form is a singlet, so the only triplet that can be built is $\langle \phi_2 \rangle$. For $\beta = 2$, the only modular form is the triplet $Y_3^{(2)}$ shown in the Appendix D. The effective triplet is the linear combination of the symmetric and antisymmetric product of the modular form with the flavon VEV, $\langle\phi_2\rangle \times Y_3^{(2)} \to 3_a + 3_s$. For $\beta = 4, 6$ the modular form can be the singlet $Y_{1'}^{(4)}, Y_1^{(6)}$ respectively and the corresponding triplets $Y_3^{(4)}, Y_{3,2}^{(6)}$, so that the actual alignment comes from the linear combination of $\langle\phi_2\rangle \times Y_{1,1'} \to 3$ and $\langle\phi_2\rangle \times Y_3 \to 3_a + 3_s$.

TABLE III. The effective alignments of the modular form $y_s^\nu$ as a triplet, depending on its weight $\alpha$. The parameter $y$ is an arbitrary real constant to comply with the extended symmetry $A_4 \ltimes \mathbb{Z}_2$.

| $\alpha$ | $(y_s^\nu)_3$ |
|---|---|
| 0 | 0 |
| 2 | $y \begin{pmatrix} 2 \\ 2\omega \\ -\omega^2 \end{pmatrix}$ |
| 4 | $y \begin{pmatrix} 2 \\ -\omega \\ 2\omega^2 \end{pmatrix}$ |
| 6 | $y \begin{pmatrix} -1 \\ 2\omega \\ 2\omega^2 \end{pmatrix}$ |

TABLE IV. The effective alignments of the modular form $y_a^\nu$ contracted with $\langle\phi_2\rangle$ into a triplet, depending on its weight $\beta$. The parameters $y_i$ are dimensionless constants constrained by the $A_4 \ltimes \mathbb{Z}_2$ symmetry.

| $\beta$ | $(y_a^\nu \langle\phi_2\rangle)_3 / v_2$ |
|---|---|
| 0 | $y_1 \begin{pmatrix} 0 \\ 1 \\ 1 \end{pmatrix}$ |
| 2 | $y_1 \begin{pmatrix} \omega^2 - 2\omega \\ -2\omega^2 - 2 \\ 4\omega - 2 \end{pmatrix} + y_2 \begin{pmatrix} -\omega^2 - 2\omega \\ -2 \\ 2 \end{pmatrix}$ |
| 4 | $y_1 \omega \begin{pmatrix} 1 \\ 0 \\ 1 \end{pmatrix} + y_2 \begin{pmatrix} -2\omega^2 + \omega \\ 4\omega^2 - 2 \\ -2\omega - 2 \end{pmatrix} + y_3 \begin{pmatrix} 2\omega^2 + \omega \\ -2 \\ 2 \end{pmatrix}$ |
| 6 | $y_1 \begin{pmatrix} 0 \\ 1 \\ 1 \end{pmatrix} + y_2 \begin{pmatrix} 2 \\ 4\omega^2 + 1 \\ 4\omega + 1 \end{pmatrix} + y_3 \begin{pmatrix} 2\omega^2 - 2\omega \\ 1 \\ -1 \end{pmatrix}$ |

By choosing the weights $\alpha$, $\beta$, the structure of the neutrino mass matrix is completely defined. In principle, the $y$ in Table III and $y_1, y_2, y_3$ in Table IV correspond to general complex numbers; however as we will see below they are constrained to comply with the nontrivial $CP$ symmetry of the model.

We have obtained all the possible $A_4$ invariant modular forms. However we have to comply with the extended symmetry $A_4 \ltimes \mathbb{Z}_2$. The $U$ generator only transforms nontrivially the triplet field $F$ which is contracted to a triplet modular form. A $U$ transformation of the field $F$ can be reabsorbed by transforming the modular form by

$$C \begin{pmatrix} 1 & 0 & 0 \\ 0 & 0 & 1 \\ 0 & 1 & 0 \end{pmatrix} \quad (33)$$

where the $C$ stands for complex conjugation. Invariant terms under the full symmetry must involve modular forms that are also invariant under the $\mathbb{Z}_2$ transformation. From Table III, the only invariant case is when $\alpha = 6$ with a real $y$. From Table IV, the only invariant cases happen when $\beta = 0$ with real $y_1$ or $\beta = 6$ with $y_{1,2}$ real and $y_3$ imaginary.

The triplet field $F$ is not only taking part in the Dirac neutrino mass terms but also in the down-quark and charged lepton mass terms; therefore they also must be invariant under the enhanced symmetry $A_4 \ltimes \mathbb{Z}_2$. In this case, the field $F$ is contracted with the flavon field $\phi_1$ and it is easy to check that the transformation in Eq. (33) leaves the VEV invariant when real and therefore the charged lepton and down-quark mass terms when the parameters $y_i^d$ and $y_i^e$ involved are real.

Finally, the modular form $y_{ij}^u$ must have weight $\alpha + 2\gamma$ to build an invariant. All the fields in the corresponding terms are singlets, so these modular forms must be singlets also and will not change the structure. Depending on $i, j$, the modular form $y_{ij}^u$ must be a different type of singlet. The weight $\alpha + 2\gamma$ has to be large enough so that the space contains the three types of singlets. This modular form does not add anything to the structure of the up-quark matrix but allows us to build the $A_4$ invariants for all $T_i T_j$ combinations. The smallest weight that allows modular forms of all three types of singlets is 20, as discussed in Appendix D. These modular forms $y_{ij}^u$ are in general complex.

The case $\beta = 0$ does not have enough freedom to fit the neutrino data. We conclude that the smallest phenomenologically viable choice for weights is

$$\alpha = \beta = 6, \quad \gamma = 7. \quad (34)$$

### E. Mass matrix structure

We are now able to express the mass matrices following Eq. (30) and the effective alignments given in Sec. III D. First, we define the dimensionless parameters

$$\langle\xi\rangle/\Lambda = \tilde{\xi} \quad \text{and} \quad v_i/\Lambda = \tilde{v}_i, \quad (35)$$

where $\Lambda$ is the original cutoff scale. The down-quark and charged lepton mass matrices are diagonal:

$$M^d = v_d \begin{pmatrix} y_1^d \tilde{\xi}^2 & 0 & 0 \\ 0 & y_2^d \tilde{\xi} & 0 \\ 0 & 0 & y_3^d \end{pmatrix} \tilde{v}_1,$$

$$M^e = v_d \begin{pmatrix} y_1^e \tilde{\xi}^2 & 0 & 0 \\ 0 & y_2^e \tilde{\xi} & 0 \\ 0 & 0 & y_3^e \end{pmatrix} \tilde{v}_1, \quad (36)$$

while the up-quark mass matrix can be written as

$$M_u = v_u \begin{pmatrix} y_{11}^u \tilde{\xi}^4 & y_{12}^u \tilde{\xi}^3 & y_{13}^u \tilde{\xi}^2 \\ y_{21}^u \tilde{\xi}^3 & y_{22}^u \tilde{\xi}^2 & y_{23}^u \tilde{\xi} \\ y_{31}^u \tilde{\xi}^2 & y_{32}^u \tilde{\xi} & y_{33}^u \end{pmatrix}, \quad (37)$$

where the parameters $y_i^d$ and $y_i^e$ are real due to the enhanced symmetry on the branes $A_4 \ltimes \mathbb{Z}_2$ while $y_{ij}^u$ are in general complex.

The down-quark and charged lepton mass matrices in Eq. (36) are diagonal so the fit to the observed masses is straightforward. The hierarchy between the masses of the different families is understood through the powers of $\tilde{\xi}$ and can be achieved assuming the dimensionless couplings to be of order $\mathcal{O}(1)$. All the contributions to quark mixing come from the up sector. The complex parameters in the up-type mass matrix [see Eq. (37)] fix the up, charm and top quark masses as well as the observed Cabibbo-Kobayashi-Maskawa (CKM) mixing angles. We can obtain a perfect fit for weight $\gamma = 7$. Different values of $\tilde{v}_1$, $\tilde{v}_2$ and $\tilde{\xi}$ can fit the observed masses using different dimensionless couplings still of order $\mathcal{O}(1)$. We show an example in Appendix E.

The form of the Dirac neutrino mass matrix depends on the weights $\alpha$ and $\beta$. All the possible alignments are given in Tables III and IV. The $\mathbb{Z}_2$ symmetry restricts ourselves to the case $\alpha = 6$ and $\beta = 0$ or $\beta = 6$. In the case of $\beta = 0$, we only have two free parameters $\{y, y_1\}$ and we cannot find a good fit with the solar and the reactor angle being too small. Therefore, the only phenomenologically viable case is for $\alpha = \beta = 6$ and we restrict ourselves to this case in the following.

As shown in Appendix B, we have to take into account the Clebsch-Gordan coefficients when contracting the modular form $(y_s^\nu F)_\mathbf{1}$ and $(y_a^\nu \langle \phi_2 \rangle F)_\mathbf{1}$ into singlets, i.e., $\mathbf{3} \times \mathbf{3} \to \mathbf{1}$, given by

$$(\varphi \psi)_\mathbf{1} = \varphi_1 \psi_1 + \varphi_2 \psi_3 + \varphi_3 \psi_2, \qquad (38)$$

after which the effective alignments for $\alpha = 6$ and $\beta = 6$ look like

$$\alpha_6 = y \begin{pmatrix} -1 \\ 2\omega^2 \\ 2\omega \end{pmatrix}, \quad \beta_6 = \begin{pmatrix} 2y_2 + y_3(2\omega^2 - 2\omega) \\ y_1 + y_2(4\omega + 1) - y_3 \\ y_1 + y_2(4\omega^2 + 1) + y_3 \end{pmatrix}, \quad (39)$$

respectively. The Dirac neutrino mass matrix is then given by

$$M_D^\nu = v_u \begin{pmatrix} (2y_2 + y_3(2\omega^2 - 2\omega))\tilde{v}_2 & -y \\ (y_1 + y_2(4\omega + 1) - y_3)\tilde{v}_2 & 2\omega^2 y \\ (y_1 + y_2(4\omega^2 + 1) + y_3)\tilde{v}_2 & 2\omega y \end{pmatrix} \tilde{\xi}. \quad (40)$$

The RH neutrino Majorana mass matrix is diagonal:

$$M_R = \langle \xi \rangle \begin{pmatrix} y_a^N \tilde{\xi}^3 & 0 \\ 0 & y_s^N \end{pmatrix}, \qquad (41)$$

with hierarchical RH neutrino masses given by the different powers of the field $\xi$. Furthermore, we have very heavy RH neutrino Majorana masses such that the left-handed neutrinos get a very small Majorana mass through type I seesaw [25]:

$$m_L^\nu = M_\nu^D M_R^{-1} (M_\nu^D)^T. \qquad (42)$$

The neutrino mass matrix looks like

$$m_\nu = \left( \frac{v_u^2}{\langle \xi \rangle} \frac{\tilde{\xi}^2}{y_s^N} \right) \alpha_6 (\alpha_6)^T + \left( \frac{v_u^2}{\langle \xi \rangle} \frac{\tilde{v}_2^2}{\tilde{\xi} y_a^N} \right) \beta_6 (\beta_6)^T, \quad (43)$$

where $\alpha_6$ and $\beta_6$ are the alignments defined in Eq. (39). The effective parameters at low energy are $\{y, y_1, y_2, y_3\}$, previously defined in Tables III and IV. The $\mathbb{Z}_2$ symmetry fixes the parameters $\{y, y_1, y_2\}$ to be real while $y_3$ is purely imaginary.

Finally we remark that this structure, with the expected hierarchy between the RH neutrinos, can give the correct baryon asymmetry of the Universe (BAU) through leptogenesis naturally. Leptogenesis is achieved through the CP violation in the neutrino Dirac mass matrix. The correct order of the BAU happens when the RH neutrino masses are $M_1 \sim 10^{10}$ GeV and $M_2 \sim 10^{13}$ GeV [26]. In this model, these are the natural expected masses as we can see from Eq. (41) and the sample fit in Appendix E. The contributions from the entries of the neutrino Dirac mass matrix and the expected BAU will fix the precise value of $M_1$. We conclude that the CP violation in the neutrino sector and the RH neutrino mass hierarchy of the model assures us that the BAU can be generated naturally [11].

### F. $\mu - \tau$ reflection symmetry

The neutrino mass matrix in Eq. (43) is $\mu - \tau$ reflection symmetric ($\mu\tau$-R symmetric). This corresponds to the interchange symmetry between the muon neutrino $\nu_\mu$ and the tau neutrino $\nu_\tau$ combined with CP symmetry, namely

$$\nu_e \to \nu_e^*, \qquad \nu_\mu \to \nu_\tau^*, \qquad \nu_\tau \to \nu_\mu^*, \qquad (44)$$

where the star superscript denotes the charge conjugation of the neutrino field. This can easily be seen from the alignments in Eq. (39) which construct the neutrino mass matrix in Eq. (43). Since the parameters $\{y, y_1, y_2\}$ are real while $y_3$ is purely imaginary, the transformation in Eq. (44) leaves the alignments invariant and accordingly the neutrino mass matrix. For a review of $\mu\tau$ symmetry see e.g., [27] and references therein; also see the recent discussion [28].

It is known that having a neutrino mass matrix $\mu\tau$-R symmetric in the flavor basis (which is our case) is equivalent to $\mu - \tau$ universal ($\mu\tau$-U) mixing in the Pontecorvo-Maki-Nakagawa-Sakata (PMNS) matrix; see Ref. [29]. The consequence of having $\mu - \tau$ symmetry is that it leads to having a nonzero reactor angle, $\theta_{13}$, together with a maximal atmospheric mixing angle and maximal Dirac CP phase:

$$\theta_{13} \neq 0, \qquad \theta_{23} = 45°, \qquad \delta^l = \pm 90°. \qquad (45)$$

We remark that this is a prediction of the model, due to having $A_4 \ltimes \mathbb{Z}_2$ symmetry on the branes.

The parameters $\{y, y_1, y_2, y_3\}$ in the neutrino mass matrix (43) will fit the rest of the PMNS observables, namely $\{\theta_{12}^l, \theta_{13}^l, \Delta m_{21}^2, \Delta m_{31}^2\}$, together with the prediction of the $\mu - \tau$ symmetry, $\theta_{23} = 45°$ and $\delta^l = -90°$. The contribution to a $\chi^2$ test function comes only from these predictions and we use the recent global fit values of neutrino data from NuFit4.0 [30]. The best fit points together with the $1\sigma$ ranges are $\theta_{23}/° = 49.6^{+1.0}_{-1.2}$ and $\delta^l/° = 215^{+40}_{-29}$ for normal mass ordering and without the Super-Kamiokande atmospheric neutrino data analysis. However, the distributions of these two observables are far from Gaussian and the predictions of having maximal atmospheric mixing angle $\theta_{23} = 45°$ and maximal $CP$ violation $\delta^l = -90°$ still lie inside the $3\sigma(4\sigma)$ region with a $\chi^2 = 5.48$ (6.81) without (with) Super-Kamiokande. Appendix E explains how a numerical fit can be performed and Table V shows two numerical fits, although this is only an example as we can find a good fit for a large range of parameters $y, y_1, y_2$ and $y_3$.[4] This is because the predictions of the model $\theta_{23} = 45°$ and $\delta^l = -90°$ are due to the $\mu\tau$-R symmetry and the four free parameters are used to fit the rest of the observables in the PMNS matrix.

TABLE V. Two examples with the four input parameters $y, y_1, y_2$ and $y_3$ that enter into the neutrino mass matrix in Eq. (43), giving the correct PMNS observables.

| Fit 1 | |
|---|---|
| Parameter | Value |
| $y$ | $-1.28$ |
| $y_1$ | $0.66$ |
| $y_2$ | $-1.05$ |
| $y_3$ | $i\,1.07$ |
| $y_s^N$ | $1$ |
| $y_a^N$ | $1$ |
| $|\tilde{\xi}|$ | $0.1$ |
| $|\tilde{v}_2|$ | $0.01$ |

| Fit 2 | |
|---|---|
| Parameter | Value |
| $y$ | $-1.00$ |
| $y_1$ | $-1.00$ |
| $y_2$ | $-0.08$ |
| $y_3$ | $i\,0.08$ |
| $y_s^N$ | $1$ |
| $y_a^N$ | $1$ |
| $|\tilde{\xi}|$ | $0.2$ |
| $|\tilde{v}_2|$ | $0.04$ |

The best fit from NuFit4.0 is for normal mass ordering with inverted ordering being disfavored with a $\Delta\chi^2 = 4.7\,(9.3)$ without (with) the Super-Kamiokande atmospheric neutrino data analysis. We tried a fit to inverted mass ordering and the $\chi^2$ test function goes up to $\chi^2 \sim 6800$. Therefore, the model predicts normal mass ordering together with maximal atmospheric mixing and $CP$ violation and a massless neutrino $m_1 = 0$ since we are only adding two RH neutrinos.

## IV. CONCLUSIONS

In this paper we have presented the first example in the literature of a GUT with a modular symmetry interpreted as a family symmetry. The theory is based on supersymmetric $SU(5)$ in 6d, where the two extra dimensions are compactified on a $T_2/\mathbb{Z}_2$ orbifold. We have shown that, if there is a finite modular symmetry, then it can only be $A_4$. Furthermore, if we restrict ourselves to the case in which $|\omega_1| = |\omega_2|$, the only possible value of the modulus parameter is $\tau = \omega = e^{i2\pi/3}$. We emphasize that this is one of the essential distinctions of the present model in contrast to recent works with modular symmetries, which regard the modulus $\tau$ as a free phenomenological parameter [15,16]. In the present paper, we assume a specific orbifold structure which fixes the modulus to a discrete choice of moduli, where we focus on the case $\tau = \omega = e^{i2\pi/3}$, although we do not address the problem of moduli stabilization.

We have shown that it is possible to construct a consistent model along these lines, which successfully combines an $SU(5)$ GUT group with the $A_4$ modular symmetry and a $U(1)$ shaping symmetry. In this model the $F$ fields on the branes are assumed to respect an enhanced symmetry $A_4 \rtimes \mathbb{Z}_2$ which leads to an effective $\mu - \tau$ reflection symmetry at low energies, which predicts the maximal atmospheric angle and maximal $CP$ phase. In addition there are two right-handed neutrinos on the branes whose Yukawa couplings are determined by modular weights, leading to specific alignments which fix the Dirac mass matrix. The model also introduces two triplet flavons in the bulk, whose vacuum alignments are determined by orbifold boundary conditions, analogous to those responsible for Higgs doublet-triplet splitting. The charged lepton and down-type quarks have diagonal and hierarchical Yukawa matrices, with quark mixing due to a hierarchical up-quark Yukawa matrix.

The resulting model, summarized in Tables I and II, provides an economical and successful description of quark and lepton (including neutrino) masses and mixing angles and $CP$ phases. Indeed the quarks can be fit perfectly, consistently with $SU(5)$, using only $\mathcal{O}(1)$ parameters. In addition we obtain a very good fit for the lepton observables with $\chi^2 \approx 5\,(7)$ without (with) Super-Kamiokande data, using four $\mathcal{O}(1)$ parameters which determine the entire lepton mixing matrix $U_{PMNS}$ and the light neutrino masses (eight observables), which implies that the theory is quite predictive. The main predictions of the model are a normal neutrino mass hierarchy with a massless neutrino, and the $\mu - \tau$ reflection symmetry predictions $\theta^l_{23} = 45°$ and $CP$ phase $\delta^l = -90°$, which will be tested soon.

---

[4]Although the model only allows the weights $\alpha = 0$ and $\beta = 0$, 6, we tried a numerical fit with all possible combinations of weights with the alignments in Tables III and IV, and the only one that worked is the $\mu\tau$-R symmetric for $\alpha = \beta = 6$.

## ACKNOWLEDGMENTS

We thank Patrick Vaudrevange for useful discussions. S. F. K. acknowledges the STFC Consolidated Grant No. ST/L000296/1 and the European Union's Horizon 2020 Research and InvisiblesPlus RISE Grant No. 690575. E. P. acknowledges the European Union's Horizon 2020 Research. S. F. K. and E. P. acknowledge the Innovation program under Marie Skłodowska-Curie grant agreement Elusives ITN Grant No. 674896.

## APPENDIX A: UNIQUENESS OF $A_4$ AS A MODULAR SYMMETRY FOR THE BRANES

In this Appendix, we show that the set of branes in Eq. (11) is invariant under the modular transformations $\bar{\Gamma}_3$ for an infinite set of discrete values of the modulus parameter $\tau$.

First, we apply the finite modular transformations in Eq. (9) on the lattice vectors

$$S\begin{pmatrix}\omega_1 \\ \omega_2\end{pmatrix} = \begin{pmatrix}\omega_2 \\ -\omega_1\end{pmatrix},$$

$$T_{(M)}\begin{pmatrix}\omega_1 \\ \omega_2\end{pmatrix} = \begin{pmatrix}e^{-2i\pi/M}\omega_1 \\ \omega_1 + e^{2i\pi/M}\omega_2\end{pmatrix}. \quad (A1)$$

Therefore, the $S$-transformed branes become

$$\bar{z}'_S = \left\{0, \frac{\omega_2}{2}, \frac{-\omega_1}{2}, \frac{\omega_2 - \omega_1}{2}\right\}. \quad (A2)$$

Using the orbifold transformations from Eq. (10), we can add $\omega_1$ to the second and fourth branes, and obtain the original set in Eq. (11). Therefore the brane set is always invariant under the $S$ transformation, for any value of $\omega_1$ and $\omega_2$.

On the other hand, the $T$-transformed branes are

$$\bar{z}'_T = \left\{0, \frac{e^{-2i\pi/M}\omega_1}{2}, \frac{\omega_1 + e^{2i\pi/M}\omega_2}{2}, \frac{e^{-2i\pi/M}\omega_1 + \omega_1 + e^{2i\pi/M}\omega_2}{2}\right\}. \quad (A3)$$

If the set is to be invariant, up to permutations of the branes, the second term in Eq. (A3) must correspond to one of the original branes.

Let us make the *Ansatz* that the second brane from Eq. (A3) corresponds to the fourth original brane $(\omega_1 + \omega_2)/2$ up to orbifold transformations, so that it must satisfy

$$e^{-i2\pi/M}\omega_1 = (2p+1)\omega_1 + (2q+1)\omega_2, \quad (A4)$$

where $p$, $q$ are general integer numbers and represent the general orbifold transformations. This relates the basis vectors $\omega_1$ and $\omega_2$ as

$$\omega_2 = \frac{e^{-i2\pi/M} - 2p - 1}{2q+1}\omega_1, \quad (A5)$$

and we may rewrite the transformed branes from Eq. (A3) in terms of $\omega_1$:

$$\bar{z}'_T = \left\{0, \frac{e^{-i2\pi/M}}{2}, \frac{1}{2} + \frac{1 - (2p+1)e^{2i\pi/M}}{4q+2}, \right.$$
$$\left. \frac{e^{-i2\pi/M}}{2} + \frac{1}{2} + \frac{1 - (2p+1)e^{2i\pi/M}}{4q+2}\right\}\omega_1. \quad (A6)$$

We now make the *Ansatz* that the third brane corresponds to the third original brane $\omega_2/2$ (and automatically the fourth corresponds to the original second one), so that it must satisfy

$$1 + \frac{1 - (2p+1)e^{2i\pi/M}}{2q+1} = 2r + (2s+1)\frac{e^{-i2\pi/M} - 2p - 1}{2q+1}, \quad (A7)$$

where $r$, $s$ are general integer numbers which represent the general orbifold transformations. After some simple manipulations this can be written as

$$(2s+1)e^{-2i\pi/M} + (2p+1)e^{2i\pi/M}$$
$$= (1-2r)(2q+1) + (2p+1)(2s+1) + 1. \quad (A8)$$

The left-hand part is complex while the right-hand side is real. To cancel the imaginary part we must have

$$s = p, \quad (A9)$$

so that

$$2(2p+1)\cos(2\pi/M)$$
$$= (1-2r)(2q+1) + (2p+1)(2p+1) + 1, \quad (A10)$$

where, since the right-hand side is an integer, the left-hand side must also be which forces

$$M = 3, 6, \quad (A11)$$

and the equation becomes

$$m(2p+1) = (1-2r)(2q+1) + (2p+1)(2p+1) + 1,$$
$$\text{where } m = (-1)^M, \quad (A12)$$

which is a Diophantine equation to be solved for $r$. We have made two *Ansätze* to obtain this equation. This is the only solution since making any other Ansatz would obtain an equation without solutions (the equation would be an odd number equal to zero). These are straightforward calculations, which are done in the same way as this one and seem repetitive to show.

The invariance condition on the branes fixes them to be:

$$\omega_2 = \frac{-e^{i2\pi/3} - 2p - 1 + (m-1)/2}{2q+1} \omega_1, \quad (A13)$$

where the choice $M = 3, 6$ only changes with $m$. Since only $\omega_2$ is physical (and not the specific integers $p, q$), we can reabsorb the $m$ dependence into $p$. As we stated before we will be studying discrete modular symmetries with $M \leq 5$ so that the only solution is $M = 3$ which fixes the relation of the branes to be

$$\omega_2 = -\frac{e^{i2\pi/3} + 2p + 2}{2q+1} \omega_1, \quad (A14)$$

where the $p$, $q$ are integers that satisfy that

$$r = \frac{(2p+1)(p+1) + q + 1}{2q+1} \quad (A15)$$

is an integer, which has infinitely many discrete solutions.

## APPENDIX B: GROUP THEORY

$A_4$ is the even permutation group of four objects, which is isomorphic to the symmetry group of a regular tetrahedron. It has 12 elements that can be generated by two generators, $S$ and $T$, with the presentation

$$S^2 = T^3 = (ST)^3 = 1. \quad (B1)$$

$A_4$ has four inequivalent irreducible representations: three singlet $\mathbf{1}, \mathbf{1}', \mathbf{1}''$ and one triplet $\mathbf{3}$ representations. We choose to work with the same complex basis as [15] and the representation matrices of the generators are shown in Table VI.

The product of two triplets, $\varphi = (\varphi_1, \varphi_2, \varphi_3)$ and $\psi = (\psi_1, \psi_2, \psi_3)$, decomposes as $\mathbf{3} \times \mathbf{3} = \mathbf{1} + \mathbf{1}' + \mathbf{1}'' + \mathbf{3}_s + \mathbf{3}_a$, where $\mathbf{3}_{s,a}$ denote the symmetric or antisymmetric product.

TABLE VI. Generators $S$ and $T$ in the irreducible representations of $A_4$, where $\omega = e^{2\pi i/3}$.

| $A_4$ | **1** | **1**' | **1**'' | **3** |
|---|---|---|---|---|
| $S$ | 1 | 1 | 1 | $\frac{1}{3}\begin{pmatrix} -1 & 2 & 2 \\ 2 & -1 & 2 \\ 2 & 2 & -1 \end{pmatrix}$ |
| $T$ | 1 | $\omega$ | $\omega^2$ | $\begin{pmatrix} 1 & 0 & 0 \\ 0 & \omega & 0 \\ 0 & 0 & \omega^2 \end{pmatrix}$ |

TABLE VII. Decomposition of the product of two triplets $\varphi, \psi$.

| | Component decomposition |
|---|---|
| $(\varphi\psi)_\mathbf{1}$ | $\varphi_1\psi_1 + \varphi_2\psi_3 + \varphi_3\psi_2$ |
| $(\varphi\psi)_{\mathbf{1}'}$ | $\varphi_3\psi_3 + \varphi_1\psi_2 + \varphi_2\psi_1$ |
| $(\varphi\psi)_{\mathbf{1}''}$ | $\varphi_2\psi_2 + \varphi_3\psi_1 + \varphi_1\psi_3$ |
| $(\varphi\psi)_{\mathbf{3}_s}$ | $\frac{1}{\sqrt{3}}\begin{pmatrix} 2\varphi_1\psi_1 - \varphi_2\psi_3 - \varphi_3\psi_2 \\ 2\varphi_3\psi_3 - \varphi_1\psi_2 - \varphi_2\psi_1 \\ 2\varphi_2\psi_2 - \varphi_3\psi_1 - \varphi_1\psi_3 \end{pmatrix}$ |
| $(\varphi\psi)_{\mathbf{3}_a}$ | $\begin{pmatrix} \varphi_2\psi_3 - \varphi_3\psi_2 \\ \varphi_1\psi_2 - \varphi_2\psi_1 \\ \varphi_3\psi_1 - \varphi_1\psi_3 \end{pmatrix}$ |

The component decompositions of the products are shown in Table VII.

The 12 elements of $A_4$ are obtained as $1, S, T, ST, TS, T^2, ST^2, STS, TST, T^2S, TST^2$ and $T^2ST$. The $A_4$ elements belong to four conjugacy classes:

$$\begin{aligned} &1C_1 : 1 \\ &4C_3 : T, ST, TS, STS \\ &4C_3^2 : T^2, ST^2, T^2S, ST^2S \\ &3C_2 : S, T^2ST, TST^2, \end{aligned} \quad (B2)$$

where $mC_n^k$ refers to the Schoenflies notation where $m$ is the number of elements of rotations by an angle $2\pi k/n$.

## APPENDIX C: GENERALIZED *CP* CONSISTENCY CONDITIONS FOR $A_4$

Here, we check the compatibility of the $\mathbb{Z}_2$ symmetry on the branes with the $A_4$ flavor symmetry. The remnant $\mathbb{Z}_2$ symmetry behaves as an effective generalized *CP* transformation and the fields on the branes will transform under $\mathbb{Z}_2$ as

$$\psi(x) \to X_\mathbf{r}\psi^*(x'), \quad (C1)$$

where $x' = (t, x_1, x_2, x_3, x_5, -x_6)$ and $X_\mathbf{r}$ is the representation matrix in the irreducible representation $r$. To combine the flavor symmetry $A_4$ with the $\mathbb{Z}_2$ symmetry, the transformations have to satisfy certain consistency conditions [31], which were specifically applied to $A_4$ flavor symmetry in [20]. These conditions assure that if we perform a $\mathbb{Z}_2$ transformation, then apply a family symmetry transformation, and finally an inverse $\mathbb{Z}_2$ transformation is followed, the resulting net transformation should be equivalent to a family symmetry transformation. It is sufficient to only impose the consistency conditions on the group generators:

$$X_\mathbf{r}\rho_\mathbf{r}^*(S)X_\mathbf{r}^{-1} = \rho_\mathbf{r}(S'), \quad X_\mathbf{r}\rho_\mathbf{r}^*(T)X_\mathbf{r}^{-1} = \rho_\mathbf{r}(T'), \quad (C2)$$

where $\rho_{\mathbf{r}}$ denotes the representation matrix for the generators $S$ and $T$; see Table VI. As shown in [20], $S'$ and $T'$ can only belong to certain conjugacy classes of $A_4$,

$$S' \in 3C_2, \qquad T' \in 4C_3 \cup 4C_3^2, \qquad (C3)$$

[see Eq. (B2) to find out the elements in each conjugacy class]. The transformations under the generalized $CP$ symmetry $\mathbb{Z}_2$ are then

$$\psi_{\mathbf{1'}} \to \psi_{\mathbf{1''}}^*, \quad \psi_{\mathbf{1''}} \to \psi_{\mathbf{1'}}^*, \quad \psi_{\mathbf{3}} \to \begin{pmatrix} 1 & 0 & 0 \\ 0 & 0 & 1 \\ 0 & 1 & 0 \end{pmatrix} \psi_{\mathbf{3}}^*, \quad (C4)$$

which are consistent with Eqs. (C2) and (C3) for $S' = S$ and $T' = T$. However in the model under consideration, we do not have any field on the branes transforming under the $\mathbf{1'}$ and $\mathbf{1''}$ representation. Thus the $\mathbb{Z}_2$ transformation only affects the $\mathbf{3}$ representations.

We conclude that the $\mathbf{3}$ representations on the brane transform under $A_4 \ltimes \mathbb{Z}_2$ as shown in Table VI and Eq. (C4).

## APPENDIX D: MODULAR FORMS

In this section, we show the construction of modular forms for $\bar{\Gamma}_3 \simeq A_4$ following [15].

In model building, the difference between the usual cited discrete symmetries, which arise as remnant symmetries of the branes, and the modular symmetries of the spacetime lattice is that, in the latter case, the fields transform, under a transformation of $\tau$, as

$$\psi \to (c\tau + d)^{-k} \rho \psi, \qquad (D1)$$

where $\rho$ is the usual matrix representation of the transformation and $k$ is called the weight and is an arbitrary number. The invariance of the action forces the usual dimensionless coupling in the superpotential $y$ to behave as [32]

$$y \to (c\tau + d)^{k_y} \rho_y y, \qquad (D2)$$

where $k_y$ is the weight and must be an even integer [24] and $\rho_y$ the usual matrix representation of the transformation. To build the invariant in global supersymmetry, we need to satisfy two conditions: first the weight $k_y$ has to cancel the overall weights of the fields and second the product of $\rho_y$ times the representation matrices of the fields has to contain an invariant singlet. When $k = 0$ for every constant, we have the usual discrete symmetry.

The weight 0 form is just a constant, singlet under $A_4$.

The first nontrivial modular form is of weight 2 that following Eq. (4) transforms as

$$Y \to (c\tau + d)^2 \rho Y. \qquad (D3)$$

The only modular forms of weight 2 behave as a triplet $Y$ and can be written as

$$Y_1(\tau) = \frac{i}{2\pi} \left( \frac{\eta'(\frac{\tau}{3})}{\eta(\frac{\tau}{3})} + \frac{\eta'(\frac{\tau+1}{3})}{\eta(\frac{\tau+1}{3})} + \frac{\eta'(\frac{\tau+2}{3})}{\eta(\frac{\tau+2}{3})} - \frac{27\eta'(3\tau)}{\eta(3\tau)} \right),$$

$$Y_2(\tau) = \frac{-i}{\pi} \left( \frac{\eta'(\frac{\tau}{3})}{\eta(\frac{\tau}{3})} + \omega^2 \frac{\eta'(\frac{\tau+1}{3})}{\eta(\frac{\tau+1}{3})} + \omega \frac{\eta'(\frac{\tau+2}{3})}{\eta(\frac{\tau+2}{3})} \right),$$

$$Y_3(\tau) = \frac{-i}{\pi} \left( \frac{\eta'(\frac{\tau}{3})}{\eta(\frac{\tau}{3})} + \omega \frac{\eta'(\frac{\tau+1}{3})}{\eta(\frac{\tau+1}{3})} + \omega^2 \frac{\eta'(\frac{\tau+2}{3})}{\eta(\frac{\tau+2}{3})} \right), \quad (D4)$$

where the $\eta(\tau)$ denotes the Dedekind function

$$\eta(\tau) = q^{1/24} \prod_{n=1}^{\infty} (1 - q^n), \qquad q \equiv e^{i2\pi\tau}. \qquad (D5)$$

There are no weight 2 singlets. In our model, the modulus field is fixed by the orbifold to be $\tau = \omega$. In this case, up to an overall coefficient, we have

$$Y_1(\omega) = 2, \qquad Y_2(\omega) = 2\omega, \qquad Y_3(\omega) = -\omega^2, \quad (D6)$$

and the triplet for weight 2 is

$$Y_{\mathbf{3}}^{(2)} = (2, 2\omega, -\omega^2). \qquad (D7)$$

Higher weight modular forms can be written in terms of the weight 2 forms by taking products of them. The weight 4 modular forms are written as

$$Y_{\mathbf{3}}^{(4)} = (Y_1^2 - Y_2 Y_3, Y_3^2 - Y_1 Y_2, Y_2^2 - Y_1 Y_3),$$
$$Y_{\mathbf{1}}^{(4)} = Y_1^2 + 2Y_2 Y_3,$$
$$Y_{\mathbf{1'}}^{(4)} = Y_3^2 + 2Y_1 Y_2,$$
$$Y_{\mathbf{1''}}^{(4)} = Y_2^2 + 2Y_1 Y_3, \qquad (D8)$$

where the subscript corresponds to the representation under $A_4$. In our model, the modulus field is fixed by the orbifold to be $\tau = \omega$. In this case, the only nonzero weight 4 modular forms are

$$Y_{\mathbf{3}}^{(4)}|_{\tau=\omega} = (2, -\omega, 2\omega^2), \qquad Y_{\mathbf{1'}}^{(4)} = \omega. \qquad (D9)$$

The weight 6 modular forms are written as

$$Y_{\mathbf{1}}^{(6)} = Y_1^3 + Y_2^3 + Y_3^3 - 3Y_1Y_2Y_3,$$
$$Y_{\mathbf{3,1}}^{(6)} = (Y_1^3 + 2Y_1Y_2Y_3, Y_1^2Y_2 + 2Y_2^2Y_3, Y_1^2Y_3 + 2Y_3^2Y_2),$$
$$Y_{\mathbf{3,2}}^{(6)} = (Y_3^3 + 2Y_1Y_2Y_3, Y_3^2Y_1 + 2Y_1^2Y_2, Y_3^2Y_2 + 2Y_2^2Y_1),$$
$$Y_{\mathbf{3,3}}^{(6)} = (Y_2^3 + 2Y_1Y_2Y_3, Y_2^2Y_3 + 2Y_3^2Y_1, Y_2^2Y_1 + 2Y_1^2Y_3).$$
(D10)

Due to relations of the Dedekind functions, the modular forms satisfy

$$Y_2^2 + 2Y_1Y_3 = 0, \quad (D11)$$

which reduce the number of possible modular forms. In our case

$$(Y_1^2 + 2Y_2Y_3)|_{\tau=\omega} = 0, \quad (D12)$$

which reduces even further the possible modular forms. The only triplet that is different from zero in Eq. (D10) is

$$Y_{\mathbf{3,2}}^{(6)}|_{\tau=\omega} = (-1, 2\omega, 2\omega^2). \quad (D13)$$

All modular forms are built from products of the weight 2 triplet. We can build the modular forms for weight 8. Following [15], this is a 15-dimensional space that must be decomposed as $2 \times \mathbf{1} + 2 \times \mathbf{1}' + 2 \times \mathbf{1}'' + 3 \times \mathbf{3}$. For simplicity we can work out only the specific case where $\tau = \omega$. This case is greatly restricted and can be checked by doing all possible multiplications of $\mathbf{3} \times \mathbf{3} \times \mathbf{3} \times \mathbf{3}$ that the only nonzero modular forms are

$$Y_{\mathbf{3}}^{(8)} = (2, 2\omega, -\omega^2), \qquad Y_{\mathbf{1}''}^{(8)} = \omega^2, \quad (D14)$$

where we can see that the triplet has the same structure as the weight 2 one. From this we conclude that any higher weight triplet would only repeat the previous structures without having any new one.

For weight 10 we would have the same triplet as in weight 4 but two singlets since we can have the nontrivial products

$$Y_{\mathbf{1}}^{(6)} \times Y_{\mathbf{1}'}^{(4)} \to \mathbf{1}', \qquad Y_{\mathbf{3}}^{(6)} \times Y_{\mathbf{3}}^{(4)} \to \mathbf{1}'', \quad (D15)$$

so that this is the first space that has two singlets. The next space that has the three singlets is built from powers of these singlets, so the modular form must have weight 20.

## APPENDIX E: NUMERICAL FIT

We perform a $\chi^2$ test function when fitting the effective neutrino mass matrix in Eq. (43) with input parameters $x = y, y_1, y_2, y_3$, from which we obtain a set of observables $P_n(x)$. We minimize the function defined as

$$\chi^2 = \sum_n \left( \frac{P_n(x) - P_n^{\text{obs}}}{\sigma_n} \right)^2, \quad (E1)$$

where the observables are given by $P_n^{\text{obs}} \in \{\theta_{12}^l, \theta_{13}^l, \theta_{23}^l, \delta^l, \Delta m_{21}^2, \Delta m_{31}^2\}$ with statistical errors $\sigma_n$. We use the recent global fit values of neutrino data from NuFit4.0 [30] and we ignore any renormalization group running corrections as well as threshold corrections associated with the two extra dimensions. Most of the observables follow an almost Gaussian distribution and we take a conservative approach using the smaller of the given uncertainties in our computations except for $\theta_{23}^l$ and $\delta^l$. The best fit from NuFit4.0 is for normal mass ordering with inverted ordering being disfavored with a $\Delta\chi^2 = 4.7\,(9.3)$ without (with) the Super-Kamiokande atmospheric neutrino data analysis. We tried a fit to inverted mass ordering and we found a $\chi^2 \sim 6800$; therefore in the following results we only focus on the case of normal mass ordering.

The model predictions are shown in Table VIII. The neutrino mass matrix in Eq. (43) predicts the maximal atmospheric mixing angle, $\theta_{23}^l = 45°$, and maximal $CP$ violation, $\delta^l = -90°$, within the $3\sigma$ region from the latest neutrino oscillation data. This is a consequence of the $\mu\tau$-R symmetric form of the neutrino mass matrix when $y, y_1, y_2$ are real while $y_3$ is imaginary. Furthermore, since we only have 2RH neutrinos, $m_1 = 0$ and there is only one physical Majorana phase $\alpha_{23}$ [33]. The predicted effective Majorana mass $m_{ee}$ [33] is also given in Table VIII.

TABLE VIII. Model predictions in the neutrino sector for weights $\alpha = \beta = 6$. The neutrino masses $m_i$ as well as the Majorana phases are pure predictions of our model. We also predict the maximal atmospheric mixing angle $\theta_{23}^l = 45°$ and maximal $CP$ phase $\delta^l = 270°$. The bound on $\sum m_i$ is taken from [34]. The bound on $m_{ee}$ is taken from [35]. There is only one physical Majorana phase since $m_1 = 0$.

| Observable | Data Central value | Data $1\sigma$ range | Model $\alpha = \beta = 6$ |
|---|---|---|---|
| $\theta_{12}^l$ (°) | 33.82 | 33.06 → 34.60 | 33.82 |
| $\theta_{13}^l$ (°) | 8.610 | 8.480 → 8.740 | 8.610 |
| $\theta_{23}^l$ (°) | 49.60 | 48.40 → 50.60 | 45.03 |
| $\delta^l$ (°) | 215.0 | 186.0 → 255.0 | 270.00 |
| $\Delta m_{21}^2/(10^{-5}\text{ eV}^2)$ | 7.390 | 7.190 → 7.600 | 7.390 |
| $\Delta m_{31}^2/(10^{-3}\text{ eV}^2)$ | 2.525 | 2.493 → 2.558 | 2.525 |
| $m_1$ (meV) | | | 0 |
| $m_2$ (meV) | | | 8.597 |
| $m_3$ (meV) | | | 50.25 |
| $\sum m_i$ (meV) | | $\lesssim 230$ | 58.85 |
| $\alpha_{23}$ (°) | | | 180.00 |
| $m_{ee}$ (meV) | | $\lesssim 60$–200 | 2.587 |

The fit has been performed using the Mixing Parameter Tools (MPT) package [36]. The values of $y, y_1, y_2$ and $y_3$ are shown in Table V. Fit 1 shows a good fit where all of the dimensionless real parameters $y$ are of $\mathcal{O}(1)$. However a large range of parameters can give an equally good fit; see for example fit 2. The VEV ratios $|\tilde{\xi}, \tilde{v}_i|$ are parameters that do not enter the fit directly and they are chosen to reproduce the hierarchy between the fermion Yukawa couplings, making the dimensionless couplings more natural numbers, i.e., $\mathcal{O}(1)$. In the case of the neutrino mass matrix, even for fixed $|\tilde{\xi}|$ and $|\tilde{v}_2|$, there is a large range of parameters $y, y_1, y_2$ and $y_3$ that can give a good fit to the observables, meaning that the modular forms for weight $\alpha = 6$ and $\beta = 6$ give a constrained form of the neutrino mass matrix which is phenomenologically suitable. For comparison, we also give the value of the $\chi^2$ test function in the case of $\beta = 0$, in which we only have two free parameters $y$ and $y_1$, and it goes up to $\chi^2 \sim 1500$, while for $\beta = 6$ with four free parameters we have found a perfect fit for a variety of values of $y, y_1, y_2$ and $y_3$.

These VEV ratios $|\tilde{\xi}, \tilde{v}_i|$ also appear in the quark and charged-lepton mass matrices in Eqs. (36) and (37). For different values of $|\tilde{\xi}|$, as in fits 1 and 2 in Table V, different dimensionless $\mathcal{O}(1)$ parameters $y_i^d, y_i^e$ and $y_{ij}^u$ can be used to give the correct mass of the down- and up-type quarks and charged leptons and we show an example in Table IX for fit 1. In this case we take into account the running of the MSSM Yukawa couplings to the GUT scale and we follow the parametrization done by [37]. The matching conditions at the SUSY scale are parametrized in terms of four parameters $\bar{\eta}_{q,b,l}$ which we set to zero and the usual $\tan\beta$ for which we choose $\tan\beta = 5$.

TABLE IX. Input parameters that enter in the charged fermion mass matrices in Eqs. (36) and (37), giving the correct charged fermion masses and CKM parameters for a choice of $\tilde{\xi}$ corresponding to fit 1. The Yukawa parameters $y_{ij}^u$ are in general complex; however most of the phases can be reabsorbed and we are left with only four physical phases $\phi_{21}, \phi_{23}, \phi_{31}$ and $\phi_{32}$ where the subscript refers to the entry in the mass matrix where we are adding the phase.

| Fit 1 | |
|---|---|
| Parameter | Value |
| $y_{11}^u$ | 1.09 |
| $y_{12}^u$ | 0.62 |
| $y_{13}^u$ | −0.44 |
| $y_{21}^u$ | −0.42 |
| $y_{22}^u$ | −0.23 |
| $y_{23}^u$ | 0.21 |
| $y_{31}^u$ | −1.24 |
| $y_{32}^u$ | −0.56 |
| $y_{33}^u$ | 0.54 |
| $\phi_{21}/°$ | 354.47 |
| $\phi_{23}/°$ | 354.67 |
| $\phi_{31}/°$ | 328.83 |
| $\phi_{32}/°$ | 329.49 |
| $|\tilde{\xi}|$ | 0.1 |
| $|\tilde{v}_1|$ | 0.01 |

| Fit 1 | |
|---|---|
| Parameter | Value |
| $y_1^d$ | 0.25 |
| $y_2^d$ | 0.49 |
| $y_3^d$ | 2.74 |
| $y_1^e$ | 0.10 |
| $y_2^e$ | 2.12 |
| $y_3^e$ | 3.60 |
| $|\tilde{\xi}|$ | 0.1 |
| $|\tilde{v}_1|$ | 0.01 |

[1] S. F. King and C. Luhn, Rep. Prog. Phys. **76**, 056201 (2013).

[2] S. F. King, Prog. Part. Nucl. Phys. **94**, 217 (2017); S. F. King, A. Merle, S. Morisi, Y. Shimizu, and M. Tanimoto, New J. Phys. **16**, 045018 (2014).

[3] G. Altarelli, F. Feruglio, and C. Hagedorn, J. High Energy Phys. 03 (2008) 052; C. Hagedorn, S. F. King, and C. Luhn, J. High Energy Phys. 06 (2010) 048; Phys. Lett. B **717**, 207 (2012); D. Meloni, J. High Energy Phys. 10 (2011) 010; B. D. Callen and R. R. Volkas, Phys. Rev. D **86**, 056007 (2012); I. K. Cooper, S. F. King, and C. Luhn, J. High Energy Phys. 06 (2012) 130; A. Meroni, S. T. Petcov, and M. Spinrath, Phys. Rev. D **86**, 113003 (2012); S. Antusch, S. F. King, and M. Spinrath, Phys. Rev. D **83**, 013005 (2011); S. F. King, C. Luhn, and A. J. Stuart, Nucl. Phys. **B867**, 203 (2013); M. Dimou, S. F. King, and C. Luhn, J. High Energy Phys. 02 (**2016**) 118; Phys. Rev. D **93**, 075026 (2016); S. Antusch, I. de Medeiros Varzielas, V. Maurer, C. Sluka, and M. Spinrath, J. High Energy Phys. 09 (2014) 141; F. Björkeroth, F. J. de Anda, I. de Medeiros Varzielas, and S. F. King, J. High Energy Phys. 06 (2015) 141; Phys. Rev. D **94**, 016006 (2016); F. Björkeroth, F. J. de Anda, S. F. King, and E. Perdomo, J. High Energy Phys. 10 (2017) 148; 12 (2017) 075.

[4] Y. Kawamura, Prog. Theor. Phys. **103**, 613 (2000); G. Altarelli and F. Feruglio, Phys. Lett. B **511**, 257 (2001); L. J. Hall and Y. Nomura, Phys. Rev. D **64**, 055003 (2001); A. Hebecker and J. March-Russell, Nucl. Phys. **B613**, 3

(2001); L. J. Hall, J. March-Russell, T. Okui, and D. Tucker-Smith, J. High Energy Phys. 09 (2004) 026; N. Haba and Y. Shimizu, Phys. Rev. D **67,** 095001 (2003); **69,** 059902(E) (2004); F. Feruglio, Eur. Phys. J. C **33,** S114 (2004); R. Dermisek and A. Mafi, Phys. Rev. D **65,** 055002 (2002); H. D. Kim and S. Raby, J. High Energy Phys. 01 (2003) 056; N. Haba, T. Kondo, and Y. Shimizu, Phys. Lett. B **535,** 271 (2002); L. J. Hall and Y. Nomura, Phys. Rev. D **66,** 075004 (2002); T. Asaka, W. Buchmuller, and L. Covi, Phys. Lett. B **523,** 199 (2001); L. J. Hall and Y. Nomura, Nucl. Phys. **B703,** 217 (2004); N. Haba and Y. Shimizu, J. High Energy Phys. 07 (2018) 057; T. Kobayashi, S. Raby, and R. J. Zhang, Phys. Lett. B **593,** 262 (2004).

[5] S. F. King and M. Malinsky, J. High Energy Phys. 11 (2006) 071.

[6] G. Altarelli, F. Feruglio, and Y. Lin, Nucl. Phys. **B775,** 31 (2007).

[7] A. Adulpravitchai and M. A. Schmidt, J. High Energy Phys. 01 (2011) 106.

[8] A. Adulpravitchai, A. Blum, and M. Lindner, J. High Energy Phys. 07 (2009) 053.

[9] T. J. Burrows and S. F. King, Nucl. Phys. **B835,** 174 (2010).

[10] T. J. Burrows and S. F. King, Nucl. Phys. **B842,** 107 (2011).

[11] F. J. de Anda and S. F. King, J. High Energy Phys. 07 (2018) 057.

[12] F. J. de Anda and S. F. King, J. High Energy Phys. 10 (2018) 128.

[13] G. Altarelli and F. Feruglio, Nucl. Phys. **B741,** 215 (2006).

[14] R. de Adelhart Toorop, F. Feruglio, and C. Hagedorn, Nucl. Phys. **B858,** 437 (2012).

[15] F. Feruglio, arXiv:1706.08749.

[16] J. C. Criado and F. Feruglio, SciPost Phys. **5,** 042 (2018); P. P. Novichkov, J. T. Penedo, S. T. Petcov, and A. V. Titov, J. High Energy Phys. 04 (2019) 005; T. Kobayashi, N. Omoto, Y. Shimizu, K. Takagi, M. Tanimoto, and T. H. Tatsuishi, J. High Energy Phys. 11 (2018) 196; J. T. Penedo and S. T. Petcov, Nucl. Phys. **B939,** 292 (2019); T. Kobayashi, K. Tanaka, and T. H. Tatsuishi, Phys. Rev. D **98,** 016004 (2018); P. P. Novichkov, J. T. Penedo, S. T. Petcov, and A. V. Titov, J. High Energy Phys. 04 (2019) 174.

[17] M. Cvetic, A. Font, L. E. Ibanez, D. Lust, and F. Quevedo, Nucl. Phys. **B361,** 194 (1991).

[18] T. Kobayashi, H. P. Nilles, F. Ploger, S. Raby, and M. Ratz, Nucl. Phys. **B768,** 135 (2007).

[19] M. Holthausen, M. Lindner, and M. A. Schmidt, J. High Energy Phys. 04 (2013) 122.

[20] G. J. Ding, S. F. King, and A. J. Stuart, J. High Energy Phys. 12 (2013) 006.

[21] C. D. Froggatt and H. B. Nielsen, Nucl. Phys. **B147,** 277 (1979).

[22] L. E. Ibanez and G. G. Ross, Phys. Lett. **110B,** 215 (1982); C.R. Phys. **8,** 1013 (2007); B. R. Greene, K. H. Kirklin, P. J. Miron, and G. G. Ross, Nucl. Phys. **B292,** 606 (1987); I. de Medeiros Varzielas, S. F. King, and G. G. Ross, Phys. Lett. B **648,** 201 (2007); S. F. King and M. Malinsky, Phys. Lett. B **645,** 351 (2007); I. de Medeiros Varzielas and G. G. Ross, arXiv:hep-ph/0612220; I. de Medeiros Varzielas, arXiv:0801.2775; R. Howl and S. F. King, Phys. Lett. B **687,** 355 (2010).

[23] S. P. Martin, Adv. Ser. Dir. High Energy Phys. **21,** 1 (2010); **18,** 1 (1998); J. E. Kim and H. P. Nilles, Phys. Lett. **138B,** 150 (1984); G. F. Giudice and A. Masiero, Phys. Lett. B **206,** 480 (1988).

[24] R. C. Gunning, *Lectures on Modular Forms* (Princeton University Press, Princeton, New Jersey, USA, 1962).

[25] P. Minkowski, Phys. Lett. **67B,** 421 (1977); T. Yanagida, in *Proceedings of the Workshop on Unified Theory and Baryon Number of the Universe*, edited by O. Sawada and A. Sugamoto (KEK, 1979) p. 95; M. Gell-Mann, P. Ramond, and R. Slansky, in *Supergravity*, edited by P. van Niewwenhuizen and D. Freedman (North Holland, Amsterdam, 1979), Conf. Proc. **C790927,** p. 315; P. Ramond, *Proceedings of the Conference C79-02-25* (1979), pp. 265–280, Report No. CALT-68-709, arXiv:hep-ph/9809459; R. N. Mohapatra and G. Senjanovic, Phys. Rev. Lett. **44,** 912 (1980); J. Schechter and J. W. F. Valle, Phys. Rev. D **22,** 2227 (1980).

[26] F. Björkeroth, F. J. de Anda, I. de Medeiros Varzielas, and S. F. King, J. High Energy Phys. 10 (2015) 104; F. Björkeroth, F. J. de Anda, I. de Medeiros Varzielas, and S. F. King, J. High Energy Phys. 01 (2017) 077.

[27] Z. z. Xing and Z. h. Zhao, Rep. Prog. Phys. **79,** 076201 (2016).

[28] S. F. King and C. C. Nishi, Phys. Lett. B **785,** 391 (2018).

[29] P. F. Harrison and W. G. Scott, Phys. Lett. B **547,** 219 (2002).

[30] I. Esteban, M. C. Gonzalez-Garcia, A. Hernandez-Cabezudo, M. Maltoni, and T. Schwetz, J. High Energy Phys. 01 (2019) 106; I. Esteban, M. C. Gonzalez-Garcia, A. Hernandez-Cabezudo, M. Maltoni, and T. Schwetz, http://www.nu-fit.org.

[31] F. Feruglio, C. Hagedorn, and R. Ziegler, J. High Energy Phys. 07 (2013) 027; W. Grimus and M. N. Rebelo, Phys. Rep. **281,** 239 (1997).

[32] S. Ferrara, D. Lust, A. D. Shapere, and S. Theisen, Phys. Lett. B **225,** 363 (1989).

[33] M. Tanabashi *et al.* (Particle Data Group), Phys. Rev. D **98,** 030001 (2018).

[34] P. A. R. Ade *et al.* (Planck Collaboration), Astron. Astrophys. **594,** A13 (2016).

[35] M. Agostini *et al.* (GERDA Collaboration), Phys. Rev. Lett. **120,** 132503 (2018); A. Gando *et al.* (KamLAND-Zen Collaboration), Phys. Rev. Lett. **117,** 082503 (2016); **117,** 109903(E) (2016).

[36] S. Antusch, J. Kersten, M. Lindner, M. Ratz, and M. A. Schmidt, J. High Energy Phys. 03 (2005) 024.

[37] S. Antusch and V. Maurer, J. High Energy Phys. 11 (2013) 115.